# Edge dependence of the Josephson current in the quantum Hall regime

Seong Jang[1], Geon-Hyoung Park[1], Kenji Watanabe[2], Takashi Taniguchi[3] and Gil-Ho Lee[1,*]

[1] Department of Physics, Pohang University of Science and Technology, Pohang, 37673, Republic of Korea,

[2] Research Center for Functional Materials, National Institute for Materials Science, Tsukuba, 305-0047, Japan

[3] International Center for Materials Nanoarchitectonics, National Institute for Materials Science, Tsukuba, 305-0047, Japan

**ABSTRACT**

The observation of Josephson current in the quantum Hall regime has attracted considerable attention, revealing the coexistence of two seemingly incompatible phases: the quantum Hall and superconducting states. However, the mechanism underlying the Josephson current remains unclear because of the observed $h/2e$ magnetic interference period and the lack of precisely quantized Hall plateaus. To address this issue, we investigate the edge dependence of the Josephson current in graphene Josephson junctions operating in the quantum Hall regime. By systematically comparing devices with native, etched, edge-free, and gate-defined edges, we demonstrate that the Josephson current is confined to the physical edges and is highly sensitive to specific edge configurations. Our findings provide direct evidence that counter-propagating quantum Hall edge states mediate Andreev bound states, enabling Josephson coupling. These results clarify the underlying mechanism of Josephson current in the quantum Hall regime and offer new strategies for engineering superconducting hybrid devices.

## I. INTRODUCTION

Hybrid systems combining superconductivity with topologically nontrivial phases offer a promising route toward realizing Majorana zero modes [1–4], a key ingredient for fault-tolerant quantum computing [5–7]. Graphene has emerged as a promising platform for this type of hybridization because of its excellent transparency in contact with various superconducting materials [8–14] and high electronic quality, enabling the establishment of robust quantum Hall (QH) states [15,16]. Several noteworthy phenomena have been reported, including the hybridization of superconductivity with integer [17–19] and fractional [20] QH states as well as Josephson coupling [21–23] in the QH regime. The first observation of Josephson current (JC) in the quantum Hall regime [21] has gained considerable attention because it demonstrates the coexistence of superconductivity which preserves time-reversal symmetry and QH states which require a strong magnetic field.

Despite these advances, the mechanism enabling Josephson coupling in the QH regime remains unresolved. A central puzzle lies in the observed magnetic interference patterns with a period of $h/2e$, contrasting with the expected $h/e$ periodicity for chiral Andreev edge states (FIG. 1(a)) [22]. Furthermore, deviations of normal-state resistance from quantized Hall plateaus suggest additional conduction channels that may facilitate JC.

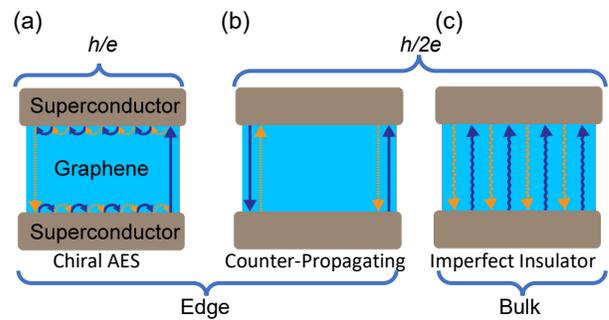

FIG. 1. Schematic representation of the three potential mechanisms for the formation of Andreev bound states in graphene Josephson junctions (GJJs) operating in the quantum Hall (QH) regime. (a) Chiral Andreev edge states (AES), (b) counter-propagating edge states, and (c) an imperfectly insulating bulk within the QH states. Blue arrows represent electron trajectories, while orange-dashed arrows denote hole trajectories, correlated through Andreev reflections.

Numerous studies have investigated the mechanism behind the observed JC in graphene Josephson junctions (GJJs) within the QH regime. The weak dependence of the JC on the GJJ width suggested that the JC was localized near the edges [24]. Furthermore, JC was shown to be switched on and off using a side gate [25], suggesting that it likely originates from the ABS along counter-propagating channels on a single edge (FIG. 1(b)). However, the observation of JC in GJJs with separations of approximately 100 nm [22,26], particularly within the plateau transition regime, suggests the possibility that JC could be mediated by the imperfectly insulating bulk of QH graphene (FIG. 1(c)). As a result, the precise mechanism responsible for JC in GJJs in the QH regime remains an open question.

To address this, we systematically investigate the role of edge configurations in mediating JC in GJJs under QH conditions. By comparing devices with native edges, etched edges, edge-free designs, and gated-defined boundaries, we identify counter-propagating edge states (CPES) as the key mechanism enabling Andreev bound state formation and Josephson coupling. Our findings clarify the origin of JC in the QH regime and provide new insights for engineering topological superconducting devices.

## II. METHODS

We fabricated GJJs with graphene encapsulated in hexagonal boron nitride (hBN) [27], which was stacked on a highly doped silicon (Si) wafer covered with a silicon dioxide ($SiO_2$) dielectric layer of a thickness of 300 nm. The device geometries and electrodes were defined using electron-beam lithography. Following plasma etching, molybdenum–rhenium (MoRe, 50%/50%) superconducting electrodes were deposited using direct current (DC) sputtering using the same electron beam resist mask to minimize contamination of the exposed graphene edges or surface. We fabricated GJJs with various edge configurations including native edge (NE), etched edge (EE), edge-free (EF), and graphite gate-defined edge (GGDE). The EE GJJs were derived from either the NE or EF GJJs for comparative analyses. The detailed edge configurations and dimensions of the junctions are described in Table S1. All measurements were performed in a dilution refrigerator (Bluefors LD400) with a nominal base temperature of 20 mK, equipped with Thermocoax cables and multistage low-pass and π-filters to minimize unwanted external electrical noise.

## III. RESULTS

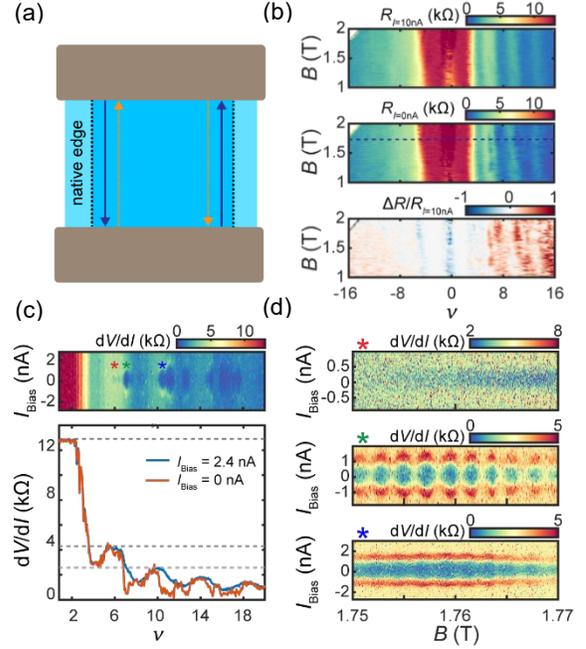

FIG. 2. Josephson current in the quantum Hall regime observed in a graphene Josephson junction (GJJ) with a native edge (NEGJJ-01). (a) Schematic of Andreev bound states formed by counter-propagating edge states in the GJJ. (b) Upper panel: fan diagram of differential resistance $R_{I=10nA}$ measured with a finite direct current (DC) bias of 10 nA and with an alternating current excitation of 100 pA. The middle panel shows a fan diagram of $R_{I=0nA}$ measured without DC bias. The lower panel shows the resistance difference $\Delta R = R_{I=10nA} - R_{I=0nA}$ normalized by $R_{I=10nA}$. (c) The upper panel shows the plot of the differential resistance ($dV/dI$) obtained as a function of bias current ($I_{Bias}$) and filling factor ($v$) at a magnetic field of $B = 1.75$ T. The lower panel shows the plot of $dV/dI$ as a function of $v$, measured with and without $I_{Bias}$. (d) $dV/dI$ as a function of $I_{Bias}$ and $B$. The upper, middle, and lower panels correspond to $v = 5.4$, 6.4, and 9.5, respectively. Colored symbols correspond to those in the upper panel of (c).

To confirm the quality of the superconducting contacts and the reliability of the measurement system, we reproduced key experimental observations from a previous study [21] using a NE GJJ (FIG. 2(a)). By comparing the differential resistance measured with and without a DC bias current ($I_{Bias}$), JC regions were visualized using red colors, as shown in the lower panel of FIG. 2(b). These regions were further investigated by varying $v$ (FIG. 2(c)) and $B$ (FIG. 2(d)). In FIG. 2(c), JC pockets appear only for $v > 5$, which does not exhibit well-quantized plateaus, suggesting the presence of additional conduction channels beyond the quantum Hall edge states, which host the ABS (FIGS. 1(b) and (c)). In FIG. 2(d), the oscillation of critical current ($I_C$) with $B$ is prominent at $v = 6.4$, barely visible at $v = 9.5$, and absent at 5.4 while the oscillation period is same as that in zero-field regime (FIG. S2). This inconsistent behavior can be explained by the CPES scenario depicted in FIG. 1(b). If both edges host JC with similar magnitudes, a SQUID-like interference pattern will emerge, as shown in the case of $v = 6.4$. If both edges host JC but with considerably different magnitudes, the interference results in small oscillation amplitudes with a finite, nonvanishing $I_C$, as shown in the case of $v = 9.5$. If one edge hosts JC while the other does not, no interference occurs, as shown in the case of $v = 5.4$.

*Corresponding author: lghman@postech.ac.kr (G.-H.L.)

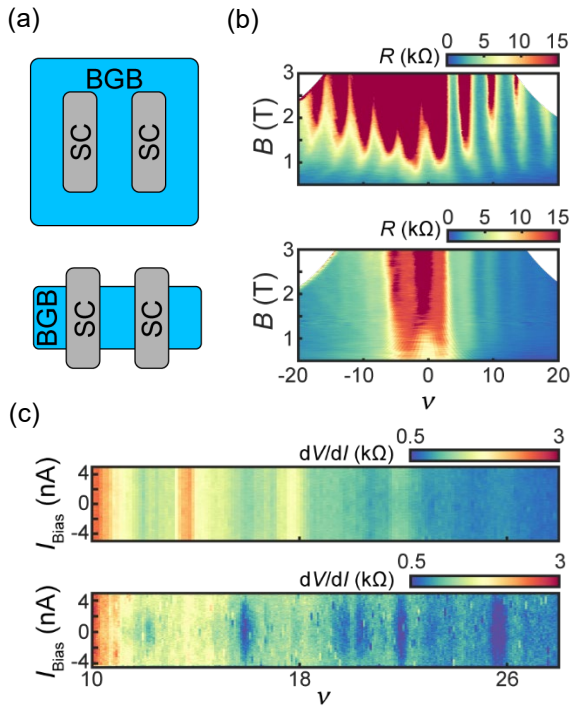

FIG. 3 Josephson current measurement in an edge-free (EF) GJJ (EFGJJ-01) and etched edge (EE) GJJ (EEGJJ-01). (a) Schematic of EF GJJ (upper panel) and EE GJJ (lower panel). The sky-blue regions represent the hBN/graphene/hBN (BGB) heterostructures. The gray regions correspond to the superconducting electrodes (SC). (b) Fan diagrams of EF GJJ (upper) and EE GJJ (lower). (c) Differential resistance as a function of $I_{Bias}$ and $v$ for EF GJJ (upper) and EE GJJ (lower) at magnetic fields of $B = 0.8$ T and 1.3 T, respectively.

To validate the CPES scenario, we examined the EF GJJ and EE GJJ devices (FIG. 3(a)) to determine whether the JC in the QH regime was mediated by the bulk of graphene. In the EF GJJ case, the resistance diverges in the QH regime, as shown in the upper panel of FIG. 3(b), where the longitudinal conductance vanishes. This indicates the absence of edge conduction channels. Consequently, no JCs were observed, excluding the possibility that bulk graphene mediates the JC when it is in the plateau transition region [22]. Upon etching, the EF GJJ transformed into the EE GJJ, exhibiting JC, as shown in FIG. 3(c), similar to the NE GJJ in FIG. 2. This abrupt change underscores the necessity of a physical edge for the JC, suggesting that ABSs reside at the edge of graphene, thus facilitating the JC.

The existence of CPES seems counterintuitive because chirality typically forbids backscattering. However, a theoretical prediction suggests that inhomogeneous screening of the electric field by charge carriers in graphene creates a nonmonotonic potential near the edge, facilitating CPES [28]. This phenomenon has been observed in several scanning probe microscopy experiments [29,30], which revealed that the CPES exists several hundred nanometers away from the physical edge. Nonetheless, recent scanning tunneling spectroscopy [31] has demonstrated that QH edge channels in graphene can be tightly confined as ideal 1D chiral states at high magnetic fields ($B$=14 T), with no evidence of electrostatic reconstruction or upstream modes. In contrast, the JC observed in our GJJ devices appears at much lower magnetic fields ($B < 2$ T), where substantial variations in the local filling factors near the edge can be present so that CPES occur.

In the presence of CPES, the ABS can be formed through Andreev reflection between the downstream and upstream mode at a single edge. However, these ABS can be disrupted by impurities near the edge, which facilitate scattering between the upstream and downstream modes and obstruct ABS formation, as illustrated in FIG. 4(a). The comparison between the NE GJJ and EE GJJ, as shown in FIG. 4, supports the notion that impurities introduced by the plasma-etching process hinder the formation of ABS [32]. To ensure a fair comparison, the NE GJJ was first measured and then etched to create an EE GJJ, allowing all comparisons to be performed on the same device. As illustrated in FIG. 4(b), the JC pocket of the EE GJJ appears notably weaker than that of the NE GJJ, and the zero-bias normalized differential resistance is higher than that of the NE GJJ. Furthermore, as shown in FIG. 4(c), the overall normalized resistance difference in the fan diagram of the NE GJJ was greater than that of the EE GJJ. These observations indicate that the JC in NE GJJ is stronger than that in EE GJJ.

This is also supported by observations from a previous scanning probe microscopy study [30], which revealed that the formation of energy levels in the antidots facilitates scattering between the upstream and downstream modes. When more impurities are introduced near the edge through the plasma etching process, the Fermi level can align more frequently, satisfying the conditions for scattering between the upstream and downstream modes. Consequently, the EE GJJ, which has more impurities near the edge compared with the NE GJJ, experiences more scattering, thereby the SC is suppressed compared with the NE GJJ.

*Corresponding author: lghman@postech.ac.kr (G.-H.L.)

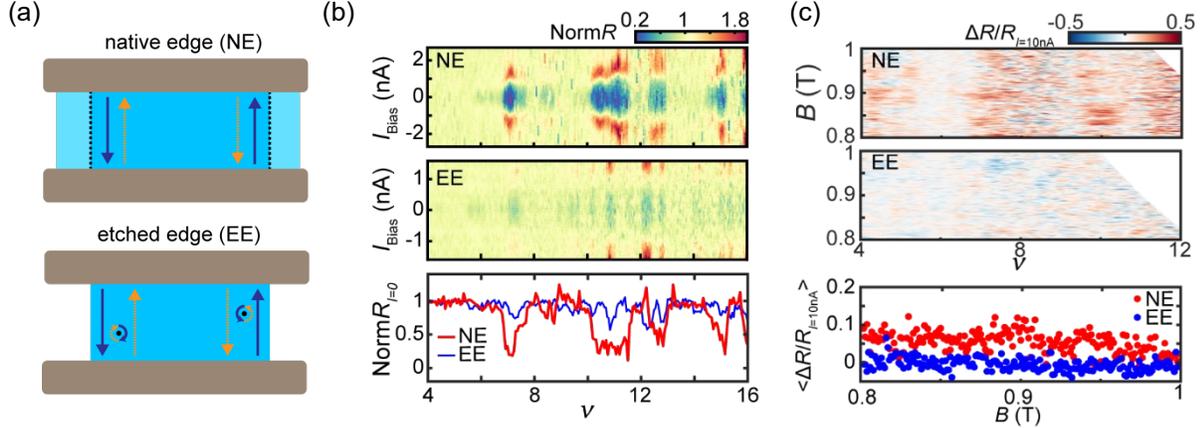

FIG. 4. (a) Schematic comparing the quantum Hall edge states and the counter-propagating edge states in a native edge GJJ (NE GJJ) and an EE GJJ. (b) Normalized differential resistance as a function of bias current ($I_{Bias}$) and filling factor ($\nu$) of NE GJJ (NEGJJ-01, upper panel) and EE GJJ (EEGJJ-02, middle panel) at a magnetic field of $B = 1.75$ T. The lower panel shows profile plots at the zero-bias current of the upper and middle panels. (c) Fan diagrams of normalized resistance difference of NE GJJ (NEGJJ-02, upper panel) and EE GJJ (EEGJJ-03, middle panel). The lower panel shows plots of the mean values of horizontal lines in the Fan diagrams as a function of the magnetic field $B$.

The formation of ABS through the Andreev coupling of the upstream and downstream modes is further supported by the strong dependence of JC in GGDE GJJ (FIG. 5(a)) on the filling factors in the local ($\nu_l$) and the surrounding global ($\nu_g$) region. The GGDE GJJ device is fabricated on hBN/graphene/hBN with a narrow graphite strip serving as the back-gate electrode to tune $\nu_l$ of graphene over the graphite (local region). Meanwhile, $\nu_g$ of graphene (global region) is controlled independently by the silicon back gate. As shown in FIG. 5(b), when $|\nu_g|$ is greater than $|\nu_l|$ and the two share the same polarity (Configuration 1), the downstream modes of the global region cancel out those of the local region, resulting in edge states with a single chirality. In this configuration, the formation of the ABS along a single edge is suppressed. In the other configurations (Configurations 2 and 3), the upstream and downstream modes form a pair of counter-propagating modes, which can form the ABS. This expectation, assuming CPES, aligns with the experimental results shown in FIGS. 5(c) and (d), where JC is absent in Configuration 1 but present in Configurations 2 and 3.

## IV. DISCUSSION

The JC pockets shown in FIG. 5(d), represented by the red color, show vertical strip patterns, meaning that JC is more sensitive to the variations in $\nu_l$ than to those in $\nu_g$. This suggests that the ABS is confined to the local region so the CPES exists only in the local region. This phenomenon is attributed to the sharp edge of the graphite gate that may induce a nonmonotonic charge concentration near the edge of the local region [28], thereby facilitating CPES formation. In contrast, the global region, where graphene is influenced by the more distant silicon gate, likely exhibits a smoother charge concentration profile at its edge, which does not support the formation of a CPES. This scenario is consistent with the conventional GJJs shown in FIGS. 2, 3, and 4, where the atomically sharp end of graphene promotes CPES formation [28].

## V. CONCLUSIONS

GJJs with various edge configurations were fabricated and analyzed to investigate the mechanisms governing ABS formation in the QH regime. All experimental observations consistently supported the CPES scenario, in which Andreev coupling between the upstream and downstream modes accounted for the observed JC in the QH regime. These findings provide a clearer understanding of how GJJs can sustain JC in the QH regime and highlight the importance of considering upstream mode when designing graphene-based QH devices.

*Corresponding author: lghman@postech.ac.kr (G.-H.L.)

Our work suggests that CPES can be further engineered to facilitate the hybridization of superconductivity with QH states by precisely controlling the edge potential, for example, through side or graphite gates. Furthermore, introducing spin polarization in CPES could provide a promising route toward realizing Majorana zero modes when combined with superconductivity.

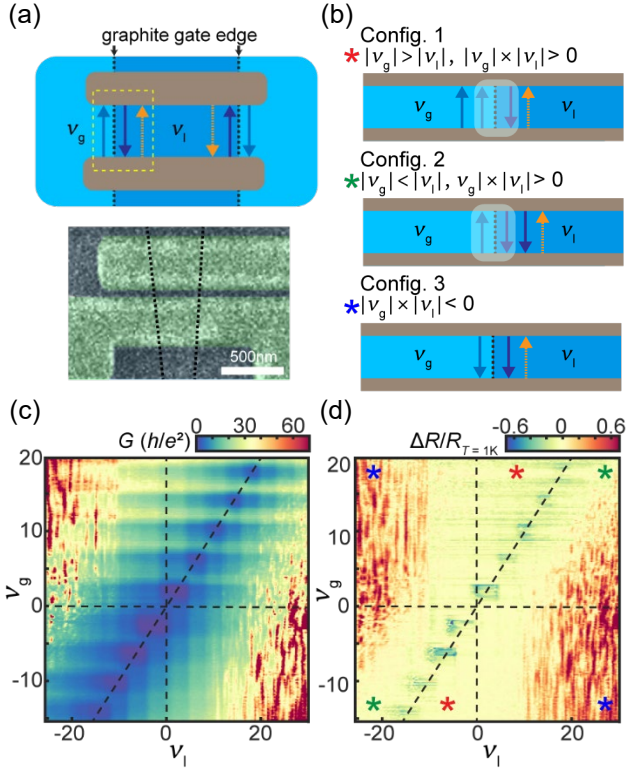

FIG. 4 (a) Upper panel shows a schematic of the graphite gate-defined edge (GGDE) GJJ. The lower panel shows a scanning electron microscopy image of GGDEGJJ-01. Dashed lines indicate the edge of the graphite gate beneath the hexagonal boron nitride (hBN)/graphene/hBN stack. The green-colored region indicates the superconducting electrodes. (b) Schematic illustrating the upstream and downstream modes of local and global region for three possible configurations, namely 1, 2, and 3. Arrows with blue, indigo, and orange colors indicate the downstream mode in the global region, that in the local region, and the upstream mode in the local region, respectively. Shading incorporating blue and indigo arrows indicates the cancellation of downstream mode in the local and global regions. (c) Conductance $G$ as a function of the filling factor for the local and global regions $v_l$ and $v_g$, respectively, measured with an AC excitation current of 1 nA at a magnetic field of $B$ = 3.1 T. (d) Normalized resistance difference obtained as a function of $v_l$ and $v_g$. Dashed lines divide the plot into three regions corresponding to the configurations illustrated in (b).

## ACKNOWLEDGMENTS

This work was supported by National Research Foundation (NRF) Grants (Nos. RS-2022-NR068223, RS-2024-00393599, RS-2024-00442710, RS-2024-00444725, 2022R1A6A3A01086903) and ITRC program (IITP-2025-RS-2022-00164799) funded by the Ministry of Science and ICT, Samsung Science and Technology Foundation (Nos. SSTF-BA2401-03 and SSTF-BA2101-06), and Samsung Electronics Co., Ltd. (IO201207-07801-01).

To whom all correspondence should be addressed: lghman@postech.ac.kr.

*Corresponding author: lghman@postech.ac.kr (G.-H.L.)

*Corresponding author: lghman@postech.ac.kr (G.-H.L.)

**Supplemental Material**

# Edge dependence of the Josephson current in the quantum Hall regime

Seong Jang[1], Geon-Hyoung Park[1], Kenji Watanabe[2], Takashi Taniguchi[3] and Gil-Ho Lee[1,*]

[1] Department of Physics, Pohang University of Science and Technology, Pohang, 37673, Republic of Korea,

[2] Research Center for Functional Materials, National Institute for Materials Science, Tsukuba, 305-0047, Japan

[3] International Center for Materials Nanoarchitectonics, National Institute for Materials Science, Tsukuba, 305-0047, Japan

## S1. Device information for graphene Josephson junction devices

| Device | Width (μm) | Length (μm) | Note |
|---|---|---|---|
| NEGJJ-01 | 1.16 | 0.3 | Native edge |
| NEGJJ-02 | 1.67 | 0.3 | Native edge |
| EFGJJ-01 | N/A | 0.3 | Edge-free |
| EEGJJ-01 | 0.9 | 0.3 | Etched from EFGJJ-01 |
| EEGJJ-02 | 0.63 | 0.3 | Etched from NEGJJ-01 |
| EEGJJ-03 | 0.7 | 0.3 | Etched from NEGJJ-02 |
| GGDEGJJ-01 | 0.28 | 0.2 | Graphite gate defined |
| GGDEGJJ-02 | 0.35 | 0.3 | Graphite gate defined |
| GGDEGJJ-03 | 0.43 | 0.35 | Graphite gate defined |

**Table S1.** Details of graphene Josephson junction devices, including device names, edge configurations, and dimensions. For graphite gate-defined Josephson junctions, the width refers to the width of the local graphite gate located beneath the graphene.

## S2. Fraunhofer pattern near zero field for NE GJJ 01

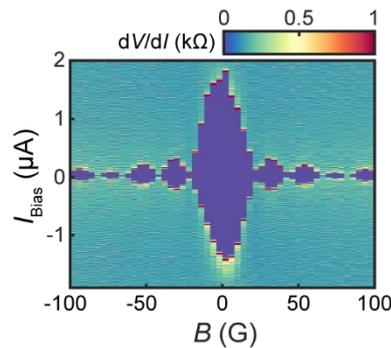

**FIG. S1**. Fraunhofer pattern of NEGJJ-01 measured at $V_{BG}$ = 20 V. The magnetic field interference period is determined to be $\Delta B$ = 20.5 G.

## S3. Additional data for graphite gate-defined edge GJJs

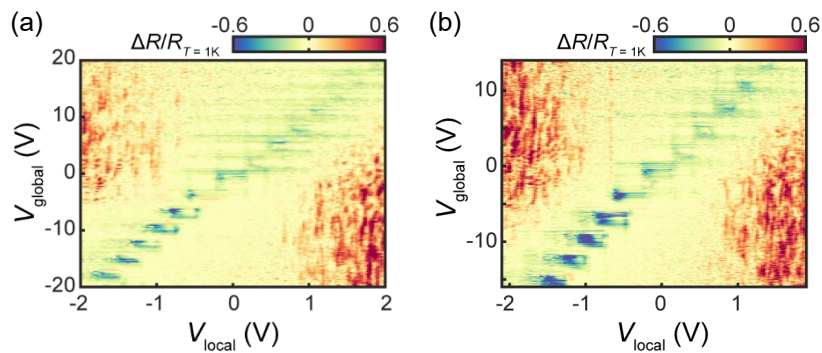

**FIG. S2.** Normalized resistance difference as a function of local graphite gate voltage ($V_{local}$) and global silicon gate voltage ($V_{global}$) for (a) GGDEGJJ-02 and (b) GGDEGJJ-03, measured with an excitation current of 1 nA at a magnetic field of $B$ = 2 T.

To ensure reproducibility, we performed the same measurement as in FIG. 5(c) and (d) on two additional graphite gate-defined graphene Josephson junction devices (GGDEGJJ-02 and GGDEGJJ-03). The results exhibit the same behavior described in FIG. 5 of the main text, where Josephson current (JC) is absent in Configuration 1 but present in Configurations 2 and 3. Additionally, the vertical striped pattern of JC pockets indicates that the Andreev bound states (ABS) are confined to the local region over the graphite gate.